\documentclass[aps,prl,twocolumn,superscriptaddress]{revtex4-1}
\usepackage{amsmath}
\usepackage{amssymb}
\usepackage{graphicx}
\usepackage{bm}

\begin{document}

\title{Doublon relaxation in the Bose-Hubbard model}

\author{A.L. Chudnovskiy}
\affiliation{1. Institut f\"ur Theoretische Physik, Universit\"at Hamburg,
Jungiusstr 9, D-20355 Hamburg, Germany}

\author{D.M.~Gangardt}
\affiliation{School of Physics and Astronomy, University of Birmingham,
Edgbaston,
Birmingham, B15 2TT, UK  }

\author{A. Kamenev}
\affiliation{Department of Physics and Fine Theoretical Physics
Institute, University of Minnesota, Minneapolis, Minnesota 55455,
USA}

\date{\today}

\begin{abstract}
Decay of a high-energy double occupancy state, doublon, in a
narrow-band lattice requires creation of a coherent many-particle
excitation. This leads to an exponentially long relaxation time of
such a state. We show that, if the average occupation number is
sufficiently small, the corresponding exponent may be evaluated
exactly. To this end we develop the quasiclassical approach to
calculation of the high-order tree-level decay amplitudes.
\end{abstract}

%\pacs{72.25.-b,75.75.+a,72.70.+m}

\maketitle

Recent experiments with cold atomic gases in optical lattices \cite{Zwerger}
allowed to create realizations of Hubbard model
%of strongly correlated systems
with  high degree of control over the main parameters, the
bandwidth $W$ and the on-site interaction $U$. In addition to
exploring rich equilibrium phase diagram \cite{Zoller98},  a rapid
time variation of the parameters  allows to prepare atomic systems
in highly excited states suitable for  studies of the far from
equilibrium dynamics of this strongly correlated system.  The most
prominent high-energy excitations are repulsively bound doubly
occupied sites called doublons \cite{Bose-Hubbard}. They were
recently observed in experiments with both bosonic
\cite{Bose-Hubbard} and fermionic \cite{DemlerPRL} atoms. The dynamical properties of doublons were considered in
recent publications \cite{DemlerPRB,Blatter10,daSilva10}.

In experiments doublon energy $U_d=U+W$ stays between the first and
second Bloch bands, which allows to consider the single band Hubbard model. In
the absence of other particles doublon is the stable excitation with a very
heavy mass \cite{Bose-Hubbard}. If other particles with the average occupation
$\rho_0$ are present, they offer a possibility for doublon to decay by
transferring its interaction energy to the kinetic energy of single-particle
excitations. In the interesting limit $\nu=U_d/W\gg 1$, the number of such
final state excited particles $n\geq \nu$ is large. Calculation of the
relaxation rate requires therefore analyzing very high orders of the
perturbation theory in the inter-particle interactions. Such an analysis for
nearly half-filled fermionic model was recently performed in
Ref.~\cite{DemlerPRB}, which found that the decay rate scales as
$\tau^{-1}\propto \exp\{-\alpha\nu\ln \nu\}$, where $\alpha$ was found to be 
approximately 0.8.

In this letter we show that if the problem admits a small parameter -- the
average filling factor $\rho_0\ll 1$, the leading exponential scaling of the
decay rate may be found exactly. We use the Bose-Hubbard version of the model
for the illustration. Our approach allows one to calculate 
generating function of {\em all}
$n$-particle tree-level threshold (i.e. such that $\nu=n$) amplitudes. Such a
generating function is shown to obey {\em classical} equation of motion for
the coupled doublon and particle fields. Luckily this equation admits an
analytic solution, which yields exact expression for the tree-level
amplitudes.  The use of the tree approximation is justified as long as $
\rho_0\nu< \gamma_D$, where $\gamma_D=e^{-3} (\pi/2e)^{D/2}$ and $D$ is spatial
dimensionality of the lattice. The similar approach was developed in
high-energy physics for calculation of the threshold amplitudes for $1\to n$
and $2\to n$ deep inelastic scattering processes \cite{Voloshin,Brown92}.

If $\rho_0 \ll \gamma_D$, the window of parameters $1\ll \nu<
\gamma_D/\rho_0$ exists where the loop diagrams may be neglected as carrying extra small factors $\rho_0$ and only tree diagrams contribute to the decay
rate.  They may be effectively summed up with the quasiclassical procedure,
described below, leading to the doublon decay rate
\begin{equation}
\frac{1}{\tau}
\propto \frac{W\nu^2}{\hbar\rho_0^2}
\,\exp\left\{- \nu\,\ln\left[\frac{\gamma_D}{\rho_0 \nu}\,
\left(\frac{2}{D}\ln \frac{\gamma_D}{\rho_0\nu}\right)^{D\over 2}\right]\right\} .
\label{decay_rate_final}
\end{equation}
The pre-exponential factor is affected by the loop diagrams and is beyond the
accuracy of the quasiclassical calculation. The factor
$[(2/D)\ln(\gamma_D/\rho_0\nu)]^{D/2}$
originates from spreading wavepacket of excited particles in $D$ spatial
dimensions \cite{CGA_unpub}. Apart from this factor, the result in
Eq.~(\ref{decay_rate_final}) reminds
Landau expression for the multi-quanta excitation of a nonlinear zero-dimensional oscillator
\cite{LandauLifshitz}.

We consider  Bose-Hubbard model on a $D$-dimensional
cubic lattice
\begin{equation}
H=-J\sum_{(i,j)} c^\dagger_i c_{j} + \frac{U}{2}\sum_{i}c^\dagger_ic^\dagger_i c_i c_i\,
\label{Bose-Hubbard}
\end{equation}
where $(i,j)$ denote nearest neighbor sites.  The first term in the
Hamiltonian, $H_0$ describes the hopping between adjacent sites
and  creates a band of the delocalized single particle states of the
width $W=4JD$  proportional to the tunneling amplitude $J$ \cite{Zoller98,Zwerger}.
Its dispersion relation is symmetric around
zero energy $\epsilon_{\bm{k}}=-2J\sum_{\mu=1}^D \cos k_{\mu}$, where $-\pi<
k_{\mu}\leq \pi$ is the quasimomentum within the first Brillouin zone,
measured in units of the inverse lattice spacing $a$. In the optical lattice,
the lattice spacing is given by the half of the laser wave length, while the
on-site repulsion $U$ is proportional  to the bulk $s$-wave scattering length
$a_s$ \cite{Zoller98,Zwerger}, which in turn can be tuned experimentally in the wide range by taking advantage of Feshbach resonance phenomenon.

The zero-temperature phase diagram  of this model
consists of the superfluid  and the Mott-insulator phases \cite{Fisher89,Zwerger}.
The Mott insulator is characterized by an
integer occupation number $\rho_0$ at each site and a finite energy gap for the
quasiparticle excitations. On the other hand, the compressible superfluid
phase is characterized by the presence of the Bose condensate with a
noninteger average occupation $\rho_0$.
We restrict ourselves with the latter scenario at $\rho_0\ll 1$.

Among the momentum content $c_{\bf k}$ of the bosonic fields  a special role
is played by the field
$c_0(t)=\sqrt{\rho_0}\, e^{-i\mu t}$, representing the condensate
with the chemical potential $\mu =-W/2$.
We shall also separate the field $c_\pi(t)$, describing quantized
particles with the energy $\epsilon_\pi-\mu=W$ at the very top of the Bloch
band. To proceed we pass to a coherent state functional representation for the
evolution operator $e^{i H t}$ and decouple the interation term in
Eq.~(\ref{Bose-Hubbard})  with the help of  the Hubbard-Stratonovich
transformation
\begin{eqnarray}
  \label{eq:HSint}
  \frac{U}{2}\bar{c}_i\bar{c}_i c_i c_i =
  \frac{\bar{d}_i d_i}{2U}-\frac{1}{2}(\bar{c}_i, c_i)\!
  \begin{pmatrix} 0 & d_i\\\bar{d}_i & 0\end{pmatrix}
  \begin{pmatrix} c_i \\ \bar{c}_i \end{pmatrix},
\end{eqnarray}
introducing the auxiliary field $d_i(t)$ which describes the doublon.
We now perform Gaussian integration over  bosonic fields
$c_{\bm k}$ with ${\bm k}\neq 0,\pi$,  using the free propagator
$G_{\bm k}(\epsilon)=(\epsilon-\epsilon_{\bm k})^{-1}$, and expand the resulting action to the second order in the $d$-fields \cite{Note2}. This leads to the doublon  Lagrangian
\begin{equation}
  \label{d-Lagrangian}
  {\cal L}_d={1\over 2} \sum_{{\bm q}}\bar d_{\bm q}(\epsilon)\big[U^{-1} -
  {\cal C}(\epsilon,{\bm q})\big] d_{\bm q}(\epsilon) \,,
\end{equation}
where ${\cal C}(\epsilon,{\bm q})$ is the Cooper polarization given by
\begin{equation}\label{Cooperon}
    {\cal C}(\epsilon,{\bm q})=
    i\!\sum_{{\bm k},\tilde\epsilon}G_{{\bm k}+{\bm
        q}}(\epsilon-\epsilon')
    G_{-{\bm k}}(\epsilon')
    = \!\sum_{\bm k}\!\frac{1}{\epsilon\!-\!\epsilon_{{\bm k}+{\bm
          q}}\!-\!\epsilon_{-{\bm k}}  }\, .
\end{equation}
The poles of the doublon propagator, i.e. ${\cal C}(\epsilon,{\bm
q})=U^{-1}$, determine its dispersion relation $\epsilon =
\epsilon_d({\bm q})$. In the limit $U\gg J$ one may expand the
denominator on the right hand side of Eq.~(\ref{Cooperon}) to the
second order to find \cite{Bose-Hubbard}
\begin{equation}\label{doublon-dispersion}
    \epsilon_d({\bm q})\approx U + \frac{8J^2}{U}\sum_{\mu=1}^D \cos^2 {q_\mu\over 2} + O(J^4/U^3)\,.
\end{equation}
Therefore the doublon band is narrow $W^2/2DU\ll W$.  One can thus
disregard the $\bm q$-dependence of the doublon dispersion and think of it as
of infinitely heavy localized particle with the energy $U$. This is equivalent
to the approximation where one takes ${\cal C}(\epsilon,{\bm q})\approx {\cal
  C}(\epsilon,\pi)=\epsilon^{-1}$. Hereafter we suppress site or momentum index of the
doublon, assuming it to be localized and infinitely heavy.

\begin{figure}
\includegraphics[width=8cm]{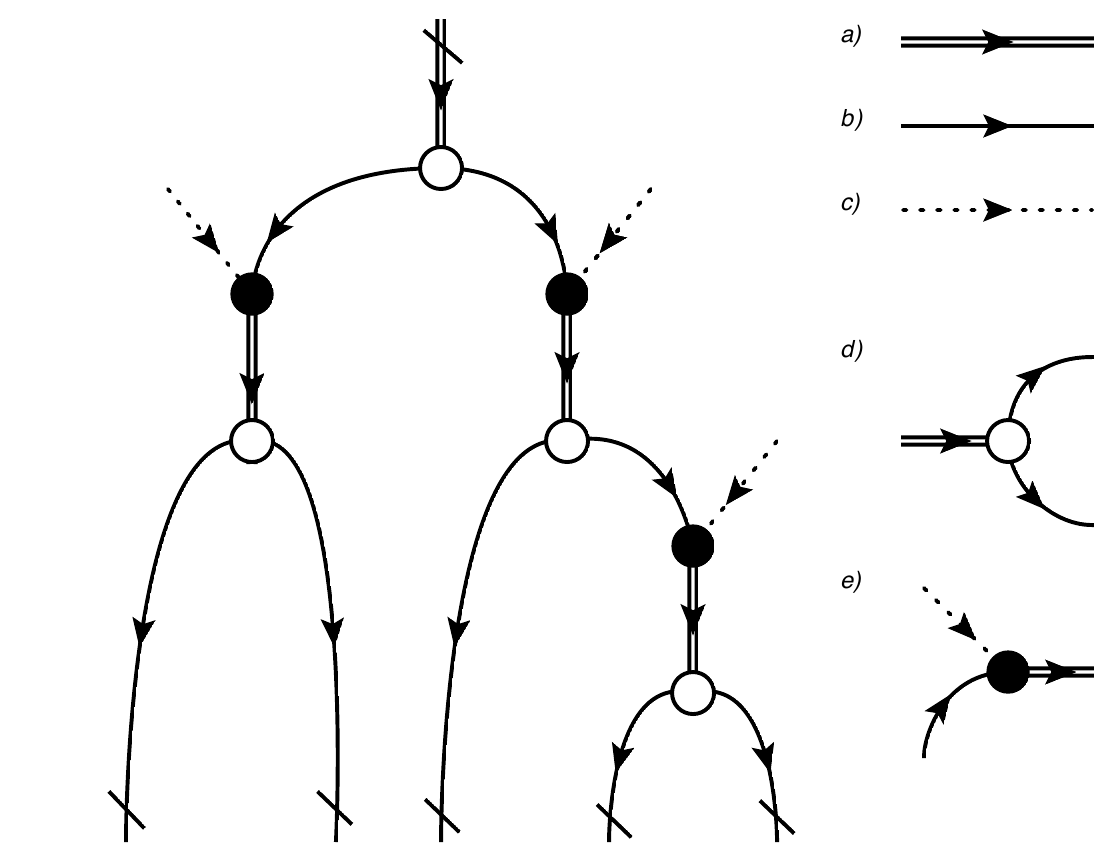}%
\caption{An example of a tree-level diagram for the amplitude of the doublon decay in $n=5$ particles.
a) Doublon propagator $\langle d \bar d\rangle$; b) particle propagator
$\langle c_\pi \bar{c}_\pi\rangle$; c) condensate amplitude
$c_0\propto \sqrt{\rho_0}$; d) doublon decay vertex $\bar{c}_\pi \bar{c}_\pi
d$; e) particle interaction with the
condensate vertex $\bar{d} c_0c_\pi$. Crossed lines denote the mass-shell particles.
\label{fig-diagram}}
\end{figure}

In the absence of the condensate, $\rho_0=0$, the doublon is absolutely stable, which is
reflected in the presence of the true pole in its Green function,
cf. Eq~(\ref{d-Lagrangian}), at $\epsilon\approx
U$. Interaction with the condensate results in the doublon self-energy
$\Sigma(\epsilon)$. Its imaginary part taken at the mass-shell is the doublon
decay rate $1/2\tau = \mathrm{Im}\Sigma(U)$, which is the focus of this
letter. With the help of the vertex $\bar{c}_\pi \bar{c}_\pi d$,
cf. Eq.~(\ref{eq:HSint}),  the doublon
initially decays onto two virtual particles with momenta close to the
boundaries of the Brillouin zone. Each of the created particles may collide
with the condensate and with the help of the vertex $\bar{d} c_\pi c_0$
create more virtual doublons, the latter again decay, {\em etc}.
This process, depicted in Fig.~\ref{fig-diagram}, continues until
the initial doublon energy (counted from the chemical
potential) $U_d=U-2\mu=U+W$ is transferred into the kinetic energy of $n\geq
U_d/W$ real particles.  The corresponding decay rate is given by the Golden
Rule
\begin{equation}
  \label{Golden-rule}
%\nonumber
  {1\over \tau} =\!\!\!\sum_{n>U_d/W} \frac{2\pi}{\hbar}\!\!
  \sum_{{\bm p}_1,\ldots, {\bm p}_{n}}\!\! \big|{\cal A}_n\big|^2\,
  \delta\!\left(U_d - \sum_{l=1}^{n}(\epsilon_{{\bm p}_l}-\mu) \!\right),
\end{equation}
$${\cal A}_n(U_d;{\bm p}_1, \ldots {\bm p}_{n})  =
\big\langle {\bm p}_1,\ldots,{\bm p}_{n}\big|\,c^\dagger_\pi c^\dagger_\pi d\,\big|d\big\rangle\,,$$
where $ \langle {\bm p}_1, \ldots {\bm p}_{n}|$ is the final $n$-particle
excited state of the {\em interacting} gas and $|d\rangle$ is the initial
state with the single doublon and rest of the gas in the condensate.

Consider now a special situation where the doublon energy happens to be
slightly below the $n$-particle creation threshold $U_d \lesssim nW$. In this
case all $n$ final particles must be very close to the top of the band. One
may thus approximate their dispersion relation as $\epsilon_{{\bm
    p}_l}-\mu=W-({\bm p}_l-\pi)^2/2m^*$, where $m^*=(2J)^{-1}$ is the
effective mass. The corresponding phase volume for the decay process
\begin{equation}\label{phase-volume}
   \! {\cal V}_n \!=\!\!\!\! \sum_{{\bm p}_1,\ldots, {\bm p}_{n}}\!\!\!\!
 \delta\!\left(\!U_d\! -\! \sum_{l=1}^{n}(\epsilon_{{\bm p}_l}\!-\!\mu)\!\! \right)\! =\frac{[(nW-U_d)/4\pi J]^{Dn\over 2}}{ \Gamma\left({Dn\over 2}\right) (nW-U_d)}
\end{equation}
depends very strongly on the deviation $nW-U_d>0$ from the threshold. On the
other hand, the corresponding matrix element ${\cal A}_n(U_d;{\bm p}_1,
\ldots {\bm p}_{n})$ appears to be a smooth function of $U_d$. This implies that
in the vicinity of the threshold it may be substituted by a constant ${\cal
  A}_n \approx {\cal A}_n(nW;\pi, \ldots  \pi)$. Indeed, once the doublon energy
is fixed at the $n$-particle threshold $U_d=nW$, all final state particles
must be at the top of the band, i.e. in the same quantum state with ${\bm
  p}=\pi$. Therefore only ${\bm p}=\pi$ state (along with the condensate ${\bm
  p}=0$) may ever appear as a real particle, while the rest of (already
integrated out) states are purely virtual. As a result the decay rate of the
doublon with the energy $U_d$ is given by
\begin{equation}
                                            \label{decay-rate}
%\nonumber
{1\over \tau(U_d)} = \frac{2\pi}{\hbar}\sum_{n>U_d/W}  \big|{\cal A}_n\big|^2\, {\cal V}_n(U_d)\, .
\end{equation}
The crack of the matter thus is to evaluate the threshold amplitudes ${\cal A}_n$.

If the doublon energy $U_d$ is not too large  the leading
contribution to the amplitude comes from the {\em tree-level} diagrams,
i.e. those which do not include loops. Indeed for the same number of the final
state mass-shell particles $n$, a loop diagram involves a higher order of the
perturbation theory than a corresponding tree one. Such tree level diagrams
are generated by repeated Wick contractions of the two types of vertices $\bar
c_\pi \bar c_\pi d$ and $\bar d c_0 c_\pi$. The Hermitian conjugated vertices
participate only in the loop diagrams. As a result, the tree level amplitudes
are generated by the following Lagrangian
\begin{equation}\label{tree-Lagrangian}
    {\cal L} = {1\over 2}\bar d\left[{1\over U} -{1\over \epsilon}\right] d +\bar c_\pi\!\left(i\partial_t\!-\!{W\over 2}\right)c_\pi - {1\over 2} \bar c_\pi \bar c_\pi d - \bar d c_\pi c_0,
\end{equation}
where the condensate field possesses a classical expectation value $c_0(t)= \sqrt{\rho_0}\, e^{iWt/2}$.
%and the factor of two in the last term is due to the fact that one of the two annihilation $c$-fields may be taken as the condensate.
Since the tree-generating Lagrangian is linear in annihilation fields
$c_\pi(t)$ and $d(t)$, the functional integration over these fields enforces
equations of motion for the corresponding creation components $\bar c_\pi(t)$
and $\bar d(t)$. To avoid the time non-local operator $1/\epsilon$ let us
define an auxiliary dimensionless field $\bar \sigma = \bar d/\epsilon$,
i.e. $\bar d(t) = -i\partial_t \bar\sigma(t)$. It is also convenient to shift
the energies for the condensate to be at zero by the gauge transformation
$\bar c_\pi\to \bar c_\pi e^{-iWt/2}$ and $\bar\sigma\to \bar\sigma e^{-iWt}$.
Taking then variation of the Lagrangian (\ref{tree-Lagrangian}) with respect
to $c_\pi$ and $d$, one obtains the following equations of motion
\begin{eqnarray}
% \nonumber to remove numbering (before each equation)
\label{c-equation}
  (-i\partial_t-W)\,\bar c_\pi &=& \sqrt{\rho_0}\,\left(-i\partial_t -W \right)\bar\sigma     \,; \\
  (-i\partial_t-U_d)\, \bar\sigma &=& U\,\bar c_\pi\bar c_\pi  \,.
  \label{sigma-equation}
\end{eqnarray}
The solution we seek may be written as a series of positive powers of $z(t) =
e^{iWt}$. For example $\bar c_\pi(t)=\sum_{n=1}^\infty \alpha_n z^n$, where
$\alpha_n$ represent the tree diagrams starting with one virtual particle and
ending up with $n$ mass-shell particles, each having the energy $W$ (thus the
factor $e^{inWt}$). The proper normalization is thus
$\alpha_1=1$.  Similarly $\bar\sigma(t)=\sum_{n=1}^\infty \beta_n z^n$,
generates tree diagrams which start from the doublon and end up with $n$
mass-shell particles, Fig.~\ref{fig-diagram}.  Since the doublon must decay at least on two particles,
the normalization is $\beta_1=0$.

The properly normalized solution of Eq.~(\ref{c-equation}) is thus
\begin{equation}
\bar c_\pi = e^{iWt}+\sqrt{\rho_0}\, \bar\sigma.
\label{c_sigma}
\end{equation}
Expressing $\bar\sigma$
and employing Eq.~(\ref{sigma-equation}), one finds
\begin{equation}\label{ricatti}
    (i\partial_t +U_d)\, \bar c_\pi + U\sqrt{\rho_0}\,\, \bar c_\pi^2 = U\, e^{iWt}\,.
\end{equation}
This equation describes a weakly non-linear high frequency $U_d$ oscillator,
which is forced with the low frequency $W\ll U_d$ driving force.  The
non-linearity generates higher harmonics of the applied force, bringing the
oscillator into the exact resonance if $U_d/W=n$ is an integer.  One expects
thus the solution to have a pole $(nW-U_d)^{-1}$, reflecting the resonance
condition. This pole represents the Green function of the incoming doublon at
the threshold energy $\epsilon=nW$. Since we are interested in the self-energy
$\Sigma(nW)$ rather than the Green function itself, we need to focus only on
the residue of the corresponding pole. The latter is a function of time given
by a series in powers of $z$. The threshold amplitude ${\cal A}_n$ is
proportional to the coefficient in front of $z^n=e^{inWt}$ term of this
series. Indeed, it is exactly the term $z^n \bar z'^n=e^{inW(t-t')}$ in
$\langle \bar d(t) d(t')\rangle$, which generates the proper delta-function in
Eq.~(\ref{Golden-rule}) upon the Fourier transform.

Equation (\ref{ricatti}) is of Riccati type, which may be transformed into the
linear second order differential equation by the substitution $\bar
c_\pi=i\partial_t v/(U\sqrt{\rho_0}\, v)$. Changing also the variable $t\to
z$ and using $\nu=U_d/W$, one finds
\begin{equation}\label{second-order-ODE}
    z\partial^2_z v - (\nu-1)\partial_z v - \sqrt{\rho_0} \,(\nu-1)^2 v=0\, .
\end{equation}
 The exact solution of
Eq. (\ref{second-order-ODE}) is given in terms of the Bessel functions
\begin{eqnarray}
\nonumber &&
v(z)=(bz)^{\nu/2}\left\{C_1 \Gamma(1-\nu) I_{-\nu}(2\sqrt{bz}) + \right. \\
&& \left.
(-1)^{\nu} C_2\Gamma(1+\nu) I_{\nu}(2\sqrt{bz})\right\},
\label{exact_solution}
\end{eqnarray}
where $b=\sqrt{\rho_0} \,(\nu-1)^2$, and $C_1$ and $C_2$ are free integration
constants. As explained above, we look for the resonant poles of the form
$z^n/(n-\nu)$ at integer values of $\nu$. It is easy to see that they may come
only from the first term on the r.h.s. of Eq.~(\ref{exact_solution}), which
reads as
\begin{equation}
v(z)= C_1\Gamma(1-\nu)\sum_{k=0}^{\infty} \frac{(bz)^{k}}{k! \Gamma(k+1-\nu)}.
\label{expansion_v}
\end{equation}
The pole structure can be made explicit by rewriting the coefficient of $z^k$
in the form
\begin{equation}
\frac{\Gamma(1-\nu)}{\Gamma(k+1-\nu)}=
\frac{1}{(k-\nu)(k-1-\nu)(k-2-\nu) \ldots (1-\nu)}.
\label{Gamma}
\end{equation}
At integer $\nu=n$ it contains poles for $k=n, n+1, \ldots $.
The resonant time dependence $z^n$ is provided only by the pole with the
lowest value $k=n$. Other pole terms contribute to the decay processes with
more than $n$ outgoing particles and are neglected.  
Notice also that at $z\to 0$ one has
$\overline{c}_{\pi}=-(W/U\sqrt{\rho_0})z\partial_z \ln v =z+O(z^2)$, which
provides the proper normalization of the generating function for tree
diagrams.
% As explained above the $n$-particle threshold amplitude ${\cal A}_n$ is
% related to the coefficient in front of $z^n/(n-\nu)$. 
The appropriate term in the
expansion of $\bar c_\pi(z)$ takes the form $[\bar c_\pi]_n z^n/(n-\nu)$,
where the coefficient
\begin{equation}
\big[\bar{c}_{\pi}\big]_n 
=(-1)^n \, \frac{(n-1)^{2n-1} \rho_0^{\frac{n-1}{2}} }{\left[(n-1)!\right]^2} \,.
\label{pole_in_c}
\end{equation}
is evaluated at $\nu=n$, 
% since near the threshold $U_d\approx nW$ and the pole
% $1/(n-\nu)$ is already explicitly separated.  
Recalling that $\bar\sigma(z) =
(\bar c_\pi-z)/\sqrt{\rho_0}$, Eq.~(\ref{c_sigma}), while the doublon field is
$\bar d(z)=Wz\partial_z\bar \sigma(z)$, one finds for the $n$-particle
threshold amplitude ${\cal A}_n=\sqrt{n!}\, [\bar d]_n$
\begin{equation}
\mathcal{A}_n=W\sqrt{n!}\, (-1)^n\, \frac{(n-1)^{2n}
\rho_0^{\frac{n}{2}-1}}{\left[(n-1)!\right]^2}\,,
\label{A_n}
\end{equation}
where the proportionality coefficient $\sqrt{n!}$ originates from the $n!$
ways of pairing $n$ final on mass-shell $\overline{c}_{\pi}$-particles in the
doublon self-energy.

The total decay rate of the doublon is calculated by summing up over all open
decay channels according to Eq.~(\ref{decay-rate}). Substituting explicit
expressions for the decay amplitude Eq. (\ref{A_n}) and the phase volume
Eq. (\ref{phase-volume}), we obtain the decay rate in form of the asymptotic
series
\begin{equation}
\frac{1}{\tau}=\frac{2\pi W}{\hbar\rho_0^2}\sum_{n>\nu} \frac{n(n-1)^{4n}
(n-\nu)^{\frac{Dn}{2}-1}}{\Gamma\left(\frac{Dn}{2}\right)\Gamma^3(n)}
\left(\frac{D^{D/2}\rho_0}{\pi^{D/2}}\right)^n.
\label{decay_rate_series}
\end{equation}
To sum up the series one employs Stirling formula for gamma
functions, substitutes summation by the integration and performs
the latter in the stationary point approximation.  As a result,
one obtains the expression (\ref{decay_rate_final}) for the
doublon decay rate. One may argue that the most general scaling
form of the decay time is given by $\ln \tau =\rho_0^{-1} h_D(x)$,
where $x = \rho_0\nu$. The dimensionless function $h_D(x)$ is
known in high energy physics as a 
``Holy Grail'' function \cite{Voloshin} . 
The semiclassical methods allow to evaluate it
in the limit $x\ll 1$, cf. Eq.~(\ref{decay_rate_final}). There are
arguments (see Ref. \cite{Voloshin}) that it either saturates, or grows
extremely slowly at $x\gtrsim 1$. Its behavior may be estimated
from comparison with experimental results reported in
Ref.~\cite{Bose-Hubbard}. The reported value of relaxation time
$\tau\approx 700 \mbox{ms}$ was measured for $\nu\approx 7.5$,
$\rho_0\approx 0.3$, and $W/\hbar \approx 1583 \mbox{Hz}$ in highly
anisotropic optical lattice corresponding to   
$D=1$. This amounts to $x\approx 2.25$.  Employing the
same functional form of the prefactor as in
Eq.~(\ref{decay_rate_final}) we estimate $h_1(2.25)\approx 4.5$.

In conclusion, the decay of doublon is accompanied by creation of a
many-particle excitation.  In a certain range of parameters, dictated by a
small filling factor, such process is described by tree-level diagrams.  If
this is the case, it is amenable to the semiclassical evaluation close in
spirit to Ref.~\cite{LandauLifshitz}.  We performed this calculation for the
Bose-Hubbard model in the superfluid phase and determined exactly
the leading terms in
the decay rate exponent.  It remains to be seen if the method and the results
can be adopted to the Fermi-Hubbard model, treated in Ref.~\cite{DemlerPRB} by
other means.

\begin{acknowledgments}

We thank A. Andreev, V. Gurarie, D.S.~Petrov and M.B. Voloshin for illuminating discussions.
We are grateful to Abdus Salam ICTP in Trieste where this project was initiated.  A.L.C. appreciates and enjoyed the hospitality  of the William I. Fine
Theoretical Physics Institute, University of Minnesota, where a part of this work has been performed. 
A.L.C.  acknowledges financial support from DFG through Sonderforschungsbereich 668. D.M.G. acknowledges support through EPSRC grant EP/D072514/1. A.K. was supported by the NSF grant  DMR-0804266.
\end{acknowledgments}

\end{document}